\documentclass[a4paper,11pt]{article}
\pdfoutput=1 % if your are submitting a pdflatex (i.e. if you have
\usepackage{jcappub} %%%HB%%%%\usepackage{jcappub} % for details on the use of the package, please
\usepackage[T1]{fontenc} % if needed

\title{\boldmath On wave dark matter in spiral and barred galaxies}

\author[a,1]{Luis A. Martinez-Medina,\note{Corresponding author. Part of the Instituto Avanzado de Cosmolog\'ia (IAC) collaboration www.iac.edu.mx}}
\author[b]{Hubert L. Bray,}
\author[a]{Tonatiuh Matos}

\affiliation[a]{Departamento de F\'isica, Centro de Investigaci\'on y de
  Estudios Avanzados del IPN, A.P. 14-740, 07000 M\'exico D.F., M\'exico.}
\affiliation[b]{Mathematics Department, Duke University, Box 90320, Durham, NC 27708, USA}

\emailAdd{lmedina@fis.cinvestav.mx}
\emailAdd{bray@math.duke.edu}
\emailAdd{tmatos@fis.cinvestav.mx}

\abstract{We recover spiral and barred spiral patterns in disk galaxy simulations with a Wave Dark Matter (WDM) background (also known as Scalar Field Dark Matter (SFDM), Ultra-Light Axion (ULA) dark matter, and Bose-Einstein Condensate (BEC) dark matter). 
Here we show how the interaction between a baryonic disk and its Dark Matter Halo triggers the formation of spiral structures when the halo is allowed to have a triaxial shape and angular momentum. This is a more realistic picture within the WDM model since a non-spherical rotating halo seems to be more natural.
By performing hydrodynamic simulations, along with earlier test particles simulations, we demonstrate another important way in which wave dark matter is consistent with observations. The common existence of bars in these simulations is particularly noteworthy.
This may have consequences when trying to obtain information about the dark matter distribution in a galaxy, the mere presence of spiral arms or a bar usually indicates that baryonic matter dominates the central region and therefore observations, like rotation curves, may not tell us what the DM distribution is at the halo center. But here we show that spiral arms and bars can develop in DM dominated galaxies with a central density core without supposing its origin on mechanisms intrinsic to the baryonic matter.}

\begin{document}
\maketitle
\flushbottom

\section{Introduction}

There are many explanations on the origin of the Spiral structure observed in disk galaxies. It is accepted that some of these spiral structures are driven by tides \citep{Dobbs2010}, by a response to bars \citep{Buta2009}, or induced by density clumps such as giant molecular clouds \citep{DOnghia2013}. 
But it is also a result of numerical simulations that spiral patterns can be self-excited structures, developing without the need of a trigger.
In general the spiral arms resulting from N-body simulations are multiple, transient features \citep{Sellwood2011, Kawata2011, Roskar2012} that span a range of pattern speeds, but there is no agreement on the lifetimes of the patterns.

Aside from the trigger mechanisms mentioned above, or the complexity of self-excited spirals seen in simulations, the interaction between a baryonic disk and its Dark Matter Halo has been studied for a few decades. It has been proposed to be responsible for a warp formation \citep{Hunter1969, Dekel1983}. A non-spherical halo misaligned with the disk imprints a torque that could form a warp.
Because dark matter halos in cosmological simulation are non-spherical, with misalignments between the principal axes of the inner and outer halo, \citet{Dubinski2009} obtain warps in their simulations when the disk is perturbed by such slowly rotating, rigid, halos.
Following this approach, it is plausible to consider a triaxial dark matter halo as the trigger mechanism of spiral structure formation in the disk.
Halos with strongly triaxial shapes are predicted by cosmological simulations within the Cold Dark Matter (CDM) scenario \citep{Allgood2006, Vera2011} via the formation process of galactic size halos occurring through the hierarchical assembly of CDM structures \citep{Klypin1999, Moore1999}.

The formation of a spiral pattern in the presence of a triaxial dark matter halo has been studied in several works. 
For example the spiral structure extended beyond optical limits in the dwarf galaxy NGC 2915 can be explained by the presence of a triaxial dark halo \citep{Bekki2002} %, Masset2003}.
A variety of halo density profiles have been used in these works, such as the NFW profile \citep{Bekki2002}, a logarithmic halo \citep{Valenzuela2014}, or a quasi-isothermal halo \citep{Khoperskov2013}. The common result of these studies is that, in spite the different dark halo models, formation of spiral structure in the stellar and gaseous disk is induced by the presence of a dark halo with triaxial shape.

The works mentioned above, that use N-body or hydrodynamics to simulate the stellar and gas components, model the DM halo as a rigid triaxial potential. By starting with a spherically symmetric DM distribution, the usual approach to account for triaxiality is to generalize  the isodensity surfaces into triaxial ellipsoids.
Here we also study the interaction between the DM halo and the disk of the galaxy by using  a rigid halo, but account for triaxiality not following the usual approach, but in a more natural way.
This paper focuses on Wave Dark Matter which we suggest as an umbrella term for 
Scalar Field Dark Matter (SFDM), Ultra-Light Axion (ULA) dark matter, and Bose-Einstein Condensate (BEC) dark matter because of the wave nature of these nearly identical models on the scale of galaxies.

Wave dark matter and the CDM model make virtually identical predictions for the distribution of dark matter at cosmological scales (\citep{Matos2001} or see recent numerical simulations confirming this result in \citep{Schive2014}). But is worth mention that the cuspy density profiles of halos in CDM-only simulations suppose a contradiction with high quality observations, a core-like behavior is preferred in the central regions of dwarfs Spheroidals (dSphs) \citep{Battaglia2008, Walker2011} and Low surface brightness galaxies (LSB) \citep{dB01, O11, A12}. CDM faces these and other difficulties on galactic scales, it appears to be necessary to look at the predictions of alternative dark matter models at scales of galaxies to overcome these difficulties. 
In our alternative scenario, dark matter is modeled by a scalar field satisfying the Klein-Gordon equation \citep{Guzman1999, Matos2000, Bernal2008, Bray2010, Marsh:2015wka}, a wave type of equation. These types of wave equations, through constructive and destructive interference, naturally form density waves which, through gravity, may induce density waves in the visible regular matter of a galaxy. Our simulations demonstrate that these visible density waves often form spiral and barred spiral patterns very similar to those observed in the universe.
Hence, the fact that wave dark matter easily produces spiral and barred spiral patterns in galaxies is very relevant to the problem of determining the model that better captures the nature of dark matter. 

The outcome that a scalar field configuration has a spacial distribution that in general is triaxial, triggering the formation of spiral and barred patterns, reinforce the hypothesis of a scalar field as dark matter. This reveals another important galaxy-like behavior of a scalar field configuration, along with the ability to model galactic systems (Lsb and dSph galaxies), as tested with observational data \citep{rob12,med14,med15}.

We comment that the recent discovery of the Higgs boson which is modeled by a scalar field as well, supports the idea that scalar fields may indeed be physical. One interpretation of the wave dark matter model is that dark matter is yet another spin zero boson, with a mass on the order of 
$10^{-22}$ eV to $10^{-20}$ eV \citep{Matos2001,BrayParry,Hlozek:2014lca} %Correct as needed
and with wavelengths perhaps approaching the scale of the smallest galaxies. In this regime, the wave-particle duality of matter causes the boson to act like a classical scalar field satisfying a wave type of equation, namely the Klein-Gordon equation.

This paper is organized as follows. First, in Sect. \ref{sec:WDM} we present
the wave dark matter triaxial potential. Then, in order to compare with another triaxial potential, in Sect. \ref{sec:NFW} we present a Navarro-Frenk-White triaxial halo. In Sect. \ref{sec:code} we briefly
describe the numerical code and initial conditions for our simulations. In Sect. \ref{sec:Results} we present the results of the simulations with the hydrodynamic code and triaxial halos. Finally, a discussion and conclusions are presented in Sect. \ref{sec:conclu}.

\section{Wave Dark Matter in a Spiral Galaxy}
\label{sec:WDM}

We model the wave dark matter in this paper precisely as in Bray (2010). We refer the reader to that paper for details and give an overview in this paper. To find approximate solutions to the nonlinear Einstein-Klein-Gordon equations, we consider exact solutions to the linear Klein-Gordon equation in a Minkowski spacetime background. We then cut these linear solutions off at an arbitrary radius. The effect of this is to produce what we claim are qualitatively realistic solutions to the nonlinear Einstein-Klein-Gordon equations, as explained in Bray (2010).

The linear solution we choose is the sum of a spherically symmetric term and a second degree spherical harmonic term. These two terms interfere, producing an interesting wave pattern which rotates around the center of the galaxy. Since we're in the Newtonian regime, we may use the Poisson equation to recover the corresponding Newtonian potential which is also rotating and is ellipsoidal. It is precisely this rotating ellipsoidal galactic potential - qualitatively predicted by the wave dark matter model - that we feed into the simulations of the stars, gas, and dust.

The addition of the second degree spherical harmonic term to the scalar field allows the wave dark matter to have angular momentum, for example, which seems natural.
We do not add in a first degree spherical harmonic term because this would cause the center of mass of the wave dark matter, and hence the regular matter, to oscillate. We believe the friction of the gas and dust of spiral galaxies would damp out this behavior. We do not add in higher degree terms because they drop off more quickly than the second degree term and are assumed to be less relevant in most situations. Future simulations could add in more terms to produce more interesting features. In particular, galaxies with three or more spiral arms could be related to instances when third degree or higher terms dominate the second degree term.

\subsection{Triaxial Potential}

As explained in Bray (2010), wave dark matter and its associated gravitational field are approximately modeled by the Poisson-Klein-Gordon equations:
\begin{eqnarray} \label{eqn:PoissonEquation}
   \Delta_x V = 4\pi \mu_{DM} &\approx& 4\pi\mu_0 \left[\left( \frac{f_t}{m} \right)^2 + f^2\right] \\
   \left( - \frac{\partial^2}{\partial t^2} + \Delta_x \right) f &=&m^2 f  \label{eqn:MKG}
\end{eqnarray}
where $V$ is the Newtonian gravitational potential of the galaxy, 
$\mu_{DM}$ is the wave dark matter density, 
$\mu_0$ is the density of the wave dark matter at the origin times the Newtonian gravitational constant $G$, 
$\Delta_x = \frac{\partial^2}{\partial x^2} + \frac{\partial^2}{\partial y^2} + \frac{\partial^2}{\partial z^2}$ is the standard Laplacian, 
$f(t,x,y,z)$ is a function (or scalar field) representing the dark matter, and
$m$ is the mass of the dark matter boson being modeled times $c/\hbar$.
Equation \ref{eqn:MKG} is a good approximation for long wavelength solutions for $f$, meaning that $f$ is changing in time much more quickly than in space. 
The main physical characteristic that we have lost in this system is that $V(t,x,y,z)$ does not feed back to affect $f(t,x,y,z)$ in the second equation.
In other words, {\it it appears} that we are not accounting for the gravity of the galaxy on the scalar field. Of course, gravity is what keeps the scalar field (and everything else) trapped in the galaxy. But we can qualitatively account for this missing effect by simply requiring that the scalar field $f$ goes to zero outside of some large radius. 

In other words, we are accounting for the effect of gravity (from all sources) on the scalar field $f$ representing dark matter, at least up to a fairly good approximation. If the total gravitational field is assumed to be spherically symmetric in the galaxy (which it roughly is), then one can study the effect this gravity has on a spherically symmetric scalar field (i.e., Bray-Goetz (2014) \citep{Bray-Goetz}, equation 10). The result is a scalar field whose spatial frequency becomes smaller as the distance from the origin increases, only to then drop off exponentially outside a given radius. However, for most of the region before the exponential drop off, the spatial frequency is quite constant, a property the solutions to equation \ref{eqn:MKG} that we'll be using have as well. While gravity will change these spatial frequencies, this is not a problem since we'll be allowed to specify our spatial frequencies as we like later. Finally, we approximate the exponential drop off by simply declaring that $f$ equals zero outside a given radius, chosen to be far outside the radius of the regular visible matter.

Solutions to Eq. \ref{eqn:MKG} can then be written as linear combinations of

\begin{equation}\label{basis}
   f = A \cos(\omega t) \cdot Y_n(\theta,\phi) \cdot r^n \cdot f_{\omega,n}(r)
\end{equation}
where
\begin{equation}
\label{sphericalode}
   f_{\omega,n}''(r) + \frac{2(n+1)}{r} f_{\omega,n}'(r) =
   (m^2 - \omega^2) f_{\omega,n}
\end{equation}
and $Y_n(\theta,\phi)$ is an $n$th degree spherical harmonic.  Note
that we require $f_{\omega,n}'(0) = 0$ but have not specified an
overall normalization.  Naturally, to get a complete basis of
solutions we also need to include solutions like the one above but
where $\cos(\omega t)$ is replaced by $\sin(\omega t)$. 

The form of the Minkowski solution on which our simulations are based is
\begin{equation}
\label{angmomsol}
   f = A_0 \cos(\omega_0 t) f_{\omega_0,0}(r) + A_2 \cos(\omega_2 t -
   2\phi) \sin^2(\theta) r^2 f_{\omega_2,2}(r)
\end{equation}
where $f_{\omega_0,0}(r)$ and $f_{\omega_2,2}(r)$ satisfy Eq.
\ref{sphericalode}. We note these solutions fall into the form of
Eq. \ref{basis} since both $\cos(2\phi)\sin^2(\theta)$ and
$\sin(2\phi)\sin^2(\theta)$ are second degree spherical harmonics.
Hence, the above Minkowski spacetime solution is the sum of a
spherically symmetric solution (degree $n=0$) and two degree two
solutions.
Note that these special solutions all have $180^\circ$ rotational symmetry with densities which rotate rigidly. Note that this rigidity is not an assumption, but a consequence of approximating away the other spherical harmonic terms, assumed to be relatively small.
As we said before, we conjecture that two armed spiral patterns result when terms similar to these dominate.

\begin{figure}
\begin{center}
\includegraphics[height=45mm]{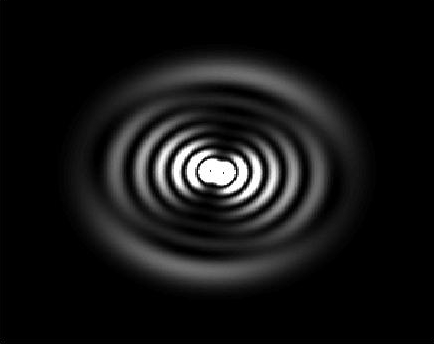} 
\hspace{-1.1cm} \includegraphics[height=45mm]{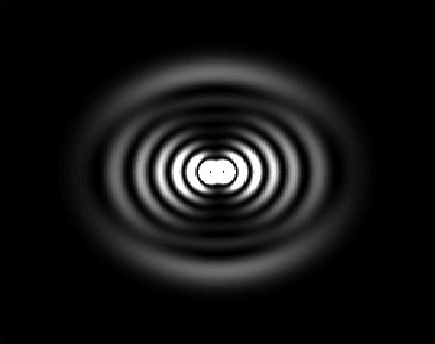} \hspace{-1.1cm} \includegraphics[height=45mm]{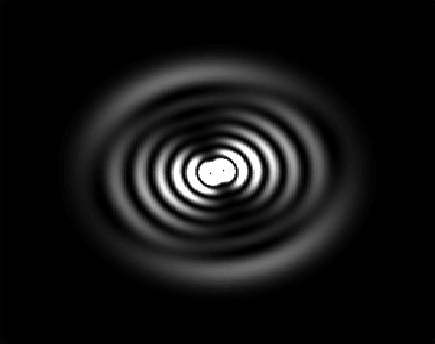}
\end{center}
\caption{Solution to the Klein-Gordon equation in a fixed
spherically symmetric potential well based on the Milky Way Galaxy
at $t = 0$, $t = 10$ Myr and $t = 20$ Myr. The
pictures show the dark matter density (in white) in the $xy$ plane.
This solution, which one can see is rotating, has angular momentum.}
\label{fig:MWG}
\end{figure}

Also, computations suggest that the approximation in
Eq. \ref{angmomsol} is a reasonable qualitative approximation
to a solution in the fixed spherically symmetric potential case. For
example, the pictures in Fig. \ref{fig:MWG} show the scalar field dark
matter densities in the $xy$ plane at $t = 0$, $t = 10$ million
years, and $t = 20$ million years in a fixed spherically symmetric
potential well based on the Milky Way Galaxy. 
Like the Minkowski
solution in Eq. \ref{angmomsol}, the solution depicted in
Fig. \ref{fig:MWG} is also a sum of a spherically symmetric solution and two
degree two solutions, but in a spherically symmetric potential.
Notice how the solutions appear to have compact support, although
actually the solution has simply decayed very rapidly outside a
finite radius. 
Otherwise, the interference pattern is very similar
to what we see with the solutions in Eq. \ref{angmomsol},
with the most obvious difference being in the last ring or two,
which are a bit stretched out in the radial direction compared to
our model. Since all we ultimately care about are the qualitative
characteristics of the gravitational potential for the dark matter
that we ultimately produce, this minor difference is something we
are willing to tolerate to get simpler equations to solve.

The reason the dark matter density is rotating in Fig. \ref{fig:MWG} and in
our Minkowski model is most easily seen by making the substitution
\begin{equation}
   \alpha = \phi - \left(\frac{\omega_2 - \omega_0}{2}\right) t
\end{equation}
into our expression for $f$ in Eq. \ref{angmomsol} to get
\begin{equation}\label{angmomsol2}
   f = A_0 \cos(\omega_0 t) f_{\omega_0,0}(r) + A_2 \cos(\omega_0 t -
   2\alpha) \sin^2(\theta) r^2 f_{\omega_2,2}(r).
\end{equation}
Note that to the extent that $\alpha$ stays fixed in time, then the
above solution does not rotate and gives a fixed interference
pattern since both terms are oscillating in time with the same
frequency $\omega_0$.  Hence,  we get
a wave dark matter interference pattern which is rotating according to
the formula
\begin{equation}\label{rotationformula}
   \phi_0 = \frac{\omega_2 - \omega_0}{2} t
\end{equation}
with period
\begin{equation}\label{periodformula}
   T_{DM} = \frac{4\pi}{\omega_2 - \omega_0},
\end{equation}
which may be related to the pattern period of the resulting barred
spiral patterns in the regular matter.

Plugging the wave function $f$ defined in Eq. \ref{angmomsol2} into equation \ref{eqn:PoissonEquation} defines both the wave dark matter density $\mu_{DM}$ in the galaxy and the corresponding galactic potential $V(t,x,y,z)$ due to the wave dark matter. Calculations using spherical harmonics as described in \citep{Bray2010} then gives us
\begin{eqnarray}\label{dmdensitysphericalharmonics}
   \frac{\mu_{DM}}{\mu_0} &\approx& U_0(r) + U_2(r) (3z^2 - r^2) +
   U_4(r) (35z^4 - 30r^2z^2 + 3r^4) \nonumber \\
   && \hspace{.4in} + \tilde{U}_2(r) (r^2 \cos(2\alpha)\sin^2(\theta))
\end{eqnarray}
where
\begin{eqnarray}
   U_0(r) &=& A_0^2 f_{\omega_0,0}(r)^2 + \frac{56}{105}A_2^2 r^4 f_{\omega_2,2}(r)^2 \nonumber \\
   U_2(r) &=& -\frac{40}{105} A_2^2 r^2 f_{\omega_2,2}(r)^2 \nonumber \\
   U_4(r) &=& \frac{3}{105} A_2^2 f_{\omega_2,2}(r)^2 \nonumber \\
   \tilde{U}_2(r) &=& 2 A_0 A_2 f_{\omega_0,0}(r) f_{\omega_2,2}(r).
   \label{sphericalharmoniccomponents}
\end{eqnarray}
Similarly, the rotating ellipsoidal galactic potential is then given by solving ode's:
\begin{eqnarray}\label{potentialsphericalharmonics}
   \frac{V}{4\pi\mu_0} &\approx& W_0(r) + W_2(r) (3z^2 - r^2) +
   W_4(r) (35z^4 - 30r^2z^2 + 3r^4) \nonumber \\
   && \hspace{.45in}+ \tilde{W}_2(r) (r^2 \cos(2\alpha)\sin^2(\theta)),
\end{eqnarray}
where for $0 \le r \le r_{max}$, 
\begin{equation}\label{potentialcomponents}
   W_n''(r) + \frac{2(n + 1)}{r} W_n'(r) = U_n(r), \;\;\;\mbox{with}\;\;\;
W_n'(0) = 0, \;\;\; W_n(r_{max}) = - \frac{r_{max}}{2n+1} W_n'(r_{max})
\end{equation}
and for $r > r_{max}$,
\begin{equation}
   W_n(r) = W_n(r_{max}) \left(\frac{r_{max}}{r}\right)^{2n+1}.
\end{equation}
Note that the above conditions are meant to apply to $\tilde{W}_2(r)$ as well.

Our input parameters for computing everything are:
\begin{itemize}
\item $\mu_0$, to be chosen to match the dark matter density in a given galaxy, 
\item $r_{max}$, the radius of the wave dark matter density,
\item $A_0$ (typically $= 1$) and $A_2$ which control the relative proportion of the  spherically symmetric and the second degree spherical harmonic components of the wave dark matter,
\item $\omega_0$ and $\omega_2$, the time frequencies of the spherically symmetric and the second degree spherical harmonic components of the wave dark matter,
\item $m$, a fundamental constant of nature in the Klein-Gordon equation, or equivalently the mass of the dark matter boson, yet to be determined precisely.
\end{itemize}

From Eq. \ref{sphericalode}, we see that the spatial wavelengths of the spherically symmetric and the second degree spherical harmonic components of the wave dark matter are
\begin{equation}
   \lambda_k = \frac{2\pi}{\sqrt{\omega_k^2 - m^2}}
\end{equation}
for $k=0,2$.  
Then
using the above equation and Eq. \ref{periodformula}, we
can solve for $\omega_0$, $\omega_2$, and $m$ in terms of
$\lambda_0$, $\lambda_2$, and $T_{DM}$ to get
\begin{eqnarray}
   \omega_0 &=& \frac{\pi}{2}\left(\frac{1}{\lambda_2^2} - \frac{1}{\lambda_0^2}\right) T_{DM} -
   \frac{2\pi}{T_{DM}}\\
   \omega_2 &=& \frac{\pi}{2}\left(\frac{1}{\lambda_2^2} - \frac{1}{\lambda_0^2}\right) T_{DM} +
   \frac{2\pi}{T_{DM}}\\
   m^2 &=& \left(\frac{2\pi}{T_{DM}}\right)^2 +
   \left(\frac{\pi T_{DM}}{2}\right)^2 \left(\frac{1}{\lambda_2^2} - \frac{1}{\lambda_0^2}\right)^2
   -2\pi^2 \left(\frac{1}{\lambda_2^2} +
   \frac{1}{\lambda_0^2}\right)\label{UpsilonEquation}.
\end{eqnarray}
Hence, if we prefer we may choose the input parameters to be 
\begin{itemize}
\item $\mu_0$, $r_{max}$, $A_0$, $A_2$, $\lambda_0$, $\lambda_2$, and $T_{DM}$,
\end{itemize}
where $T_{DM}$ is the time it takes the ellipsoidal galactic potential to rotate around once. These input parameters are more convenient in many instances and are probably more closely related to observable properties of galaxies.

\begin{figure}
   \begin{center}
   \includegraphics[height=61mm]{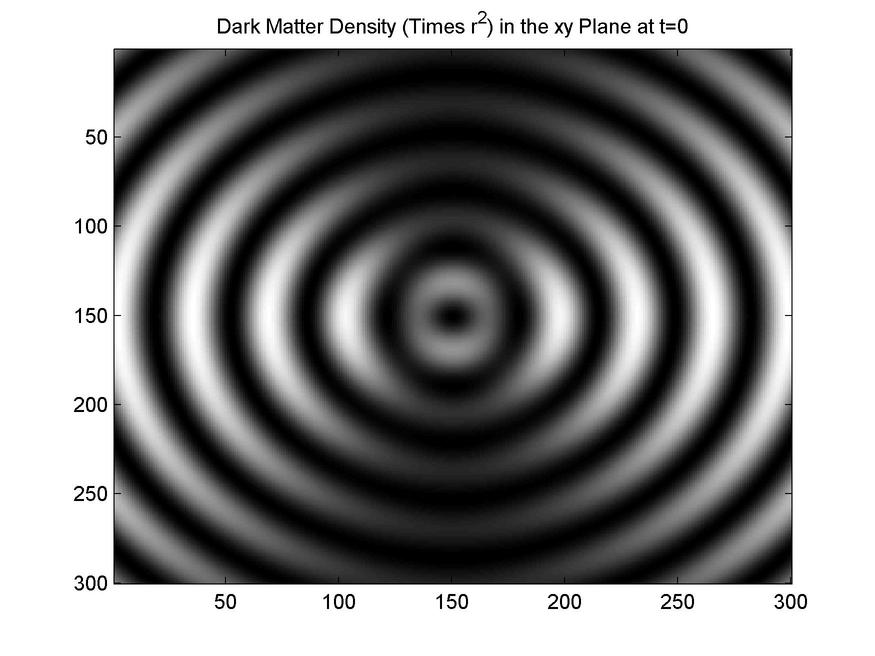}
   \hspace{-1.2cm}
   \includegraphics[height=61mm]{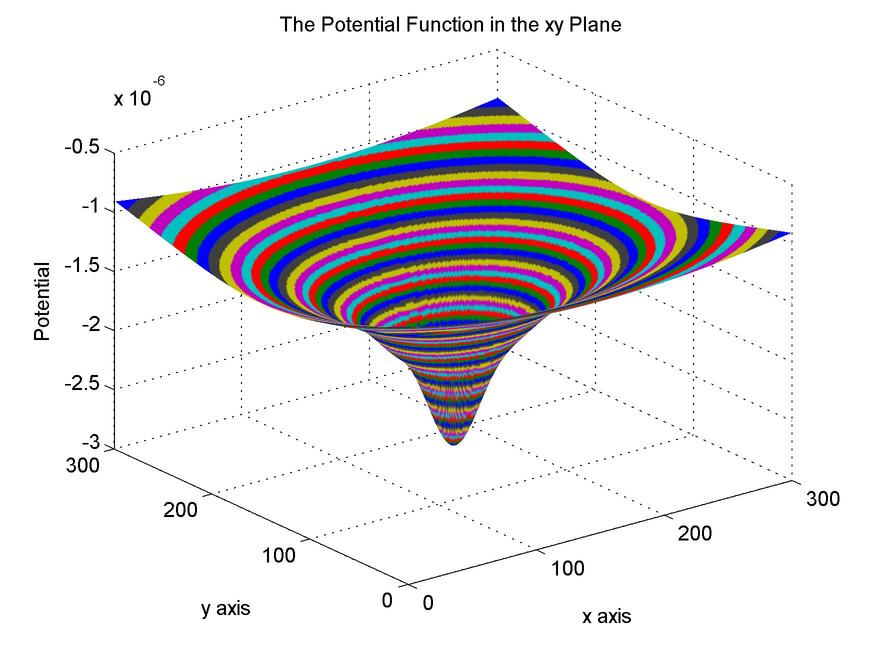}
   \includegraphics[height=61mm]{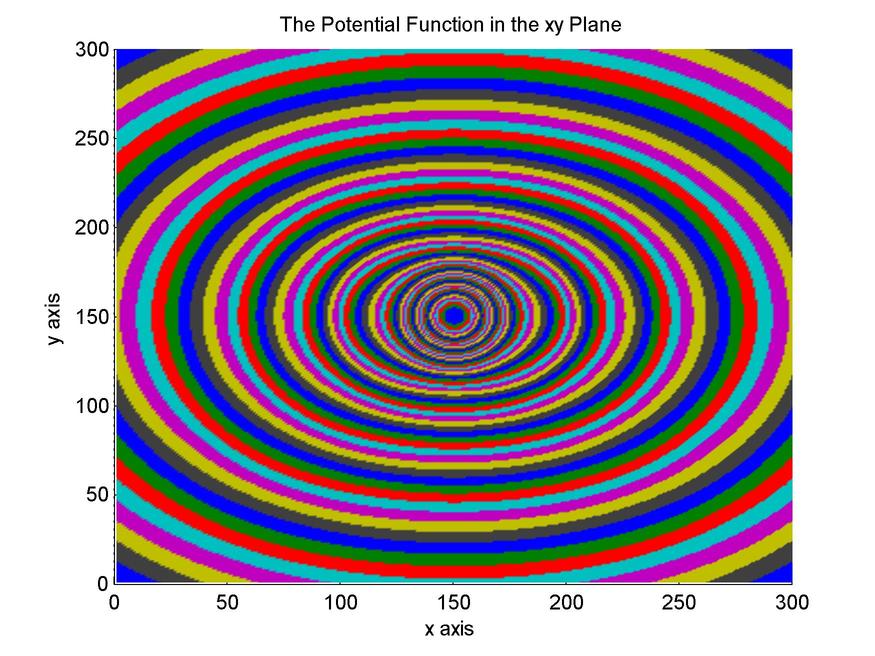}
      \hspace{-1.2cm}
   \includegraphics[height=61mm]{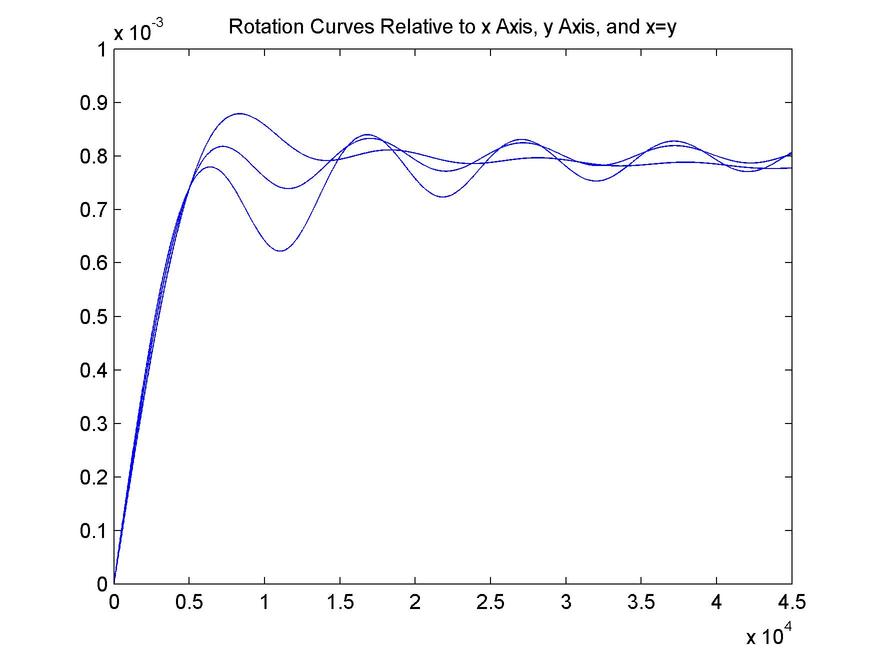}
   \end{center}
   \caption{Spiral Galaxy Simulation \#4 in \cite{Bray2010}:  The dark matter density times $r^2$ in the $xy$ plane,
   the potential function in the $xy$ plane, the level sets of the potential function
   in the $xy$ plane, and the rotation curves. \label{fig:data4}}
\end{figure}

Bray (2010), in a series of test particles simulations, use the triaxial potential of Eq. \ref{potentialsphericalharmonics} to see the imprints that the wave dark matter, modeled in this way, produce on a disk of particles. Fig. \ref{fig:data4}, originally from that paper, shows how interference patterns in the wave dark matter may result in a rotating ellipsoidal galactic potential with plausible rotation curves.

\section{Comparison with a NFW triaxial halo}
\label{sec:NFW}

As mentioned in the Introduction, there are in the literature some works dealing with the effects that a DM halo, with triaxial shape, has on the disk of its galaxy. For that reason here we also use a NFW halo to compare our results with the already existing studies.

A NFW halo results from CDM simulations and is characterized by a density profile that rise steeply towards the center \citep{N10}

\begin{equation}
\label{eq:NFW}
\rho_{NFW}(r) = \frac{\rho_s}{(r/r_s)(1+r/rs)^2},
\end{equation}
with $\rho_s$ and $r_s$ as fitting parameters.

Given a spherically symmetric density distribution $\rho(r)$, Chandrasekhar (1969) \citep{Chandrasekhar1969}  gives a detailed procedure to generalize it into a triaxial distribution. Knowing $\rho(r)$, the gravitational potential due to the generalized halo with triaxial shape is given by

\begin{equation}
\Phi = \pi G\rho a_1a_2a_3 \int_{0}^{\infty}\frac{du}{\Delta}\int_{m^2(u)}^1dm^2\rho(m^2),
\label{eq:ElipsoideHetero}
\end{equation}
where $a_1$, $a_2$ and $a_3$ are the semi axes of the triaxial distribution, and the function $m^2(u)$ (not to be confused with the scalar field mass $m$) is defined as

\begin{equation}
m^2(u) = \sum_{i=1}^3\frac{x_i^2}{a_i^2 + u} ;
\end{equation}

By putting $\rho_{NFW}$ inside Eq. \ref{eq:ElipsoideHetero}, the gravitational potential at an inner point of a NFW triaxial halo is given by

\begin{eqnarray}
  \Phi_{NFW}(x,y,z)=2\pi G abc\rho_0 r_s^2 \hspace{3cm} \nonumber \\
 \times \int_0^\infty \frac{m(u)/r_s}{1+m(u)/r_s} \frac{du } {\sqrt{(a^2+u)(b^2+u)(c^2+u)}}.
\label{TriNFW}
\end{eqnarray}

Finally we put this triaxial halo to rotate as a rigid body.

We have two halo models with triaxial shapes accounted by very different approaches. It is also important to mention that although both kind of halos are triaxial, the dark matter nature proposed by each model (CDM and WDM) has implications on the density profiles of the halos, mainly on the inner regions. A NFW halo has a cuspy behavior whereas a WDM halo has a core behavior. So although both allow triaxiality, this will differentiate the models by using observational data, for example rotation curves.

\section{The code}
\label{sec:code}

The simulations shown in this work were carried out using the latest version of the ZeusMP code \citep{H06} to model the baryonic component of a galaxy as a fluid which, as most of the material in an astrophysical system, has zero viscosity, very high compressibility, and low densities.

By contrast, in \citep{Bray2010} the stars, gas, and dust were modeled using test particles. One big difference with a fluid model is the requirement that there is a flow velocity vector $\vec u$ at each point which makes velocity a function of position, which is not the case with n-body simulations. However, in actual spiral galaxies, stars, gas, and dust do have very similar velocities when they are close to each other, which favors modeling them as a fluid in this regard. Hence, we consider the simulations presented here to be an important improvement over the previous simulations, plus being fully three-dimensional.

With this hydrodynamic code the dark matter is implemented as an external gravitational potential in the gas momentum equation. This means that besides being affected by thermal pressure gradients, a fluid element in our simulations will feel the external influence of the gravitational potential in Eq. \ref{potentialsphericalharmonics} or Eq. \ref{TriNFW}.

For each hydrodynamic simulation we set up the initial condition of a three-dimensional gas distribution as an isolated, cold, Miyamoto-Nagai disk with a density profile

\begin{equation}
\label{eq:densityprofile}
\rho_d = \frac{b^2M}{4\pi}\frac{aR^2+(a+3\sqrt{z^2+b^2})(a+\sqrt{z^2+b^2})^2}{(R^2+(a+\sqrt{z^2+b^2})^2)^{5/2}(z^2+b^2)^{3/2}}.
\end{equation}
where $M$ is the mass of the galaxy disk, $a$ and $b$ are the radial and vertical scale-length, respectively.

We model the fluid as an ideal gas with internal energy density constant trough all the disk at the initial time. Fitting parameters to rotation curves data for LSB galaxies prefer thin disks, hence we chose a small value for the internal energy density ($~0.006 J/m^3$), which means that the system is supported mainly by rotation, i.e., for the initial condition the gas component is distributed in the velocity space in such a way that it is supported by rotation against the background gravitational force due to the dark matter halo. In the next section we will see that the gas distribution does not undergoes a significant evolution other than a response to the triaxiality of the halo, with spiral patterns correlated to the semi axes of the distribution.

\section{Results}
\label{sec:Results}

First we test the potential models in Eqs. \ref{potentialsphericalharmonics} and \ref{TriNFW} against rotation curves data.
As the main goal is to compare the two dark halos, we pick data from low surface brightness galaxies (LSB). An important feature of LSB galaxies is that the mass-to-luminosity ratio is usually higher than that of a normal spiral galaxy due to the low luminosity, and the dark matter fraction is much higher, which allows to model these systems as a unique component, a dark matter halo, neglecting the gravitational influence of the luminous matter. 

\begin{figure}
   \begin{center}
   \includegraphics[height=60mm]{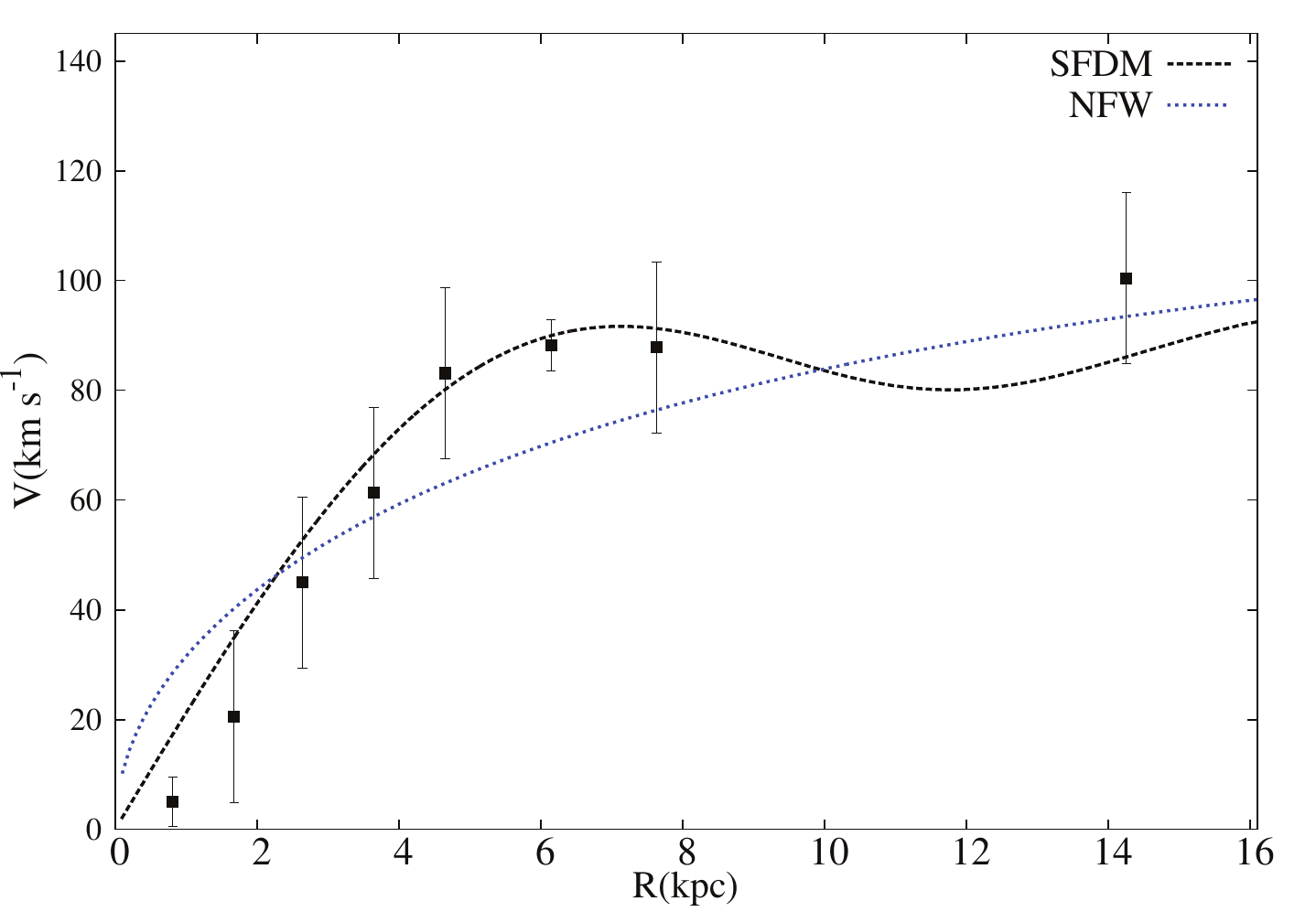}
   \end{center}
   \caption{Initial rotation curve for the two triaxial halos adjusted to the data of the LSB galaxy F568-3.}
   \label{fig:F5683}
\end{figure}

We set the parameters of the two models to fit the rotation curve data of HI in the LSB galaxy F568-3. This allow us to stablish the initial values in the simulations but we do not intend to model a particular galaxy. As we can see in Fig.  \ref{fig:F5683}, the rotation curve profiles are very different, specially in the inner parts, this reveals the cuspy nature of the NFW halo and the core nature of the WDM halo.

With this initial fit we set the parameters of the two models (Eqs. \ref{potentialsphericalharmonics} and \ref{TriNFW}) and evolve the hydrodynamics equations of a gas disk embedded in each triaxial halo. 

\begin{figure*}
   \begin{center}
   \hspace{-1.3cm}   \includegraphics[height=80mm]{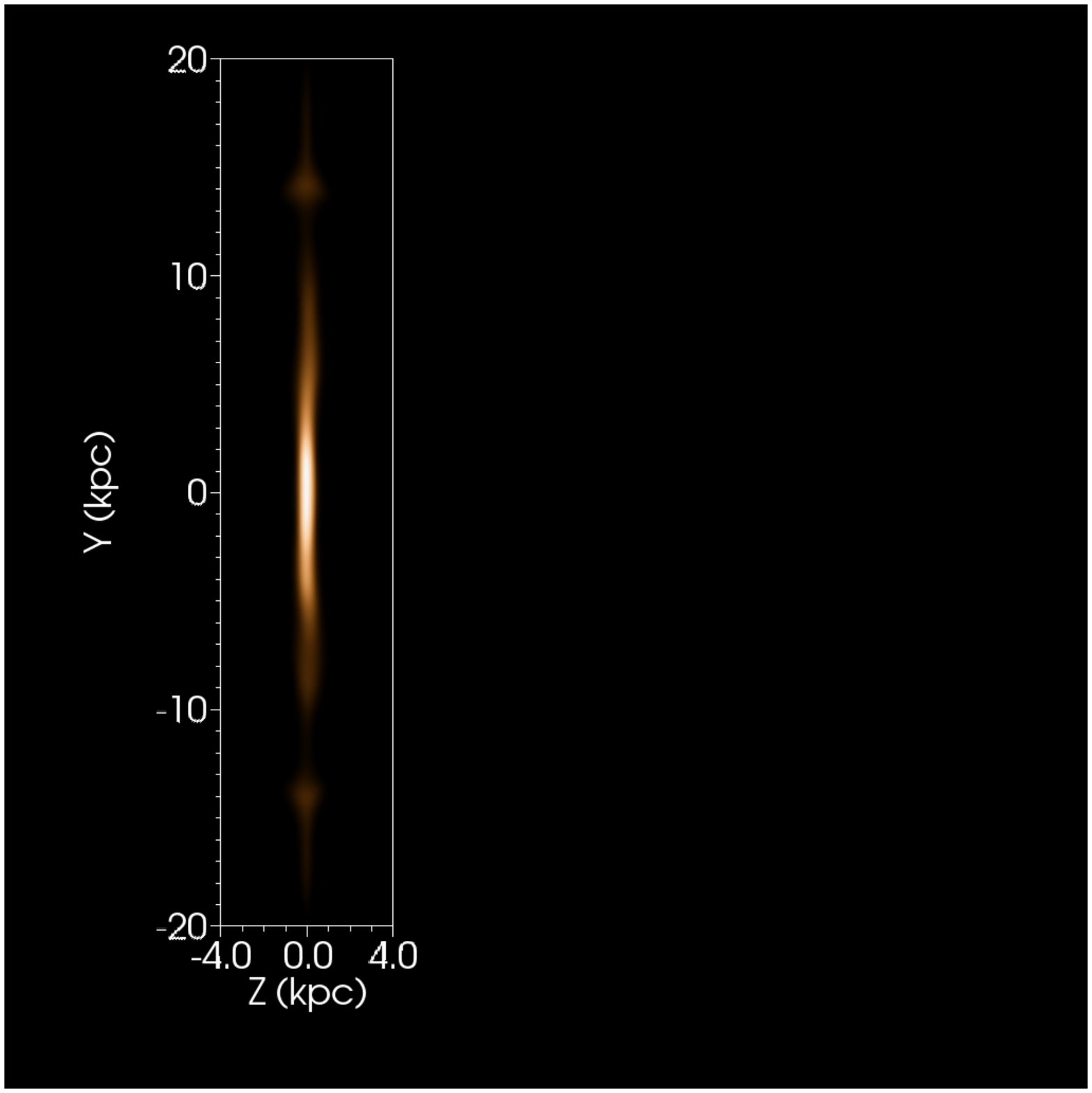} 
   \hspace{-4.3cm} \includegraphics[height=80mm]{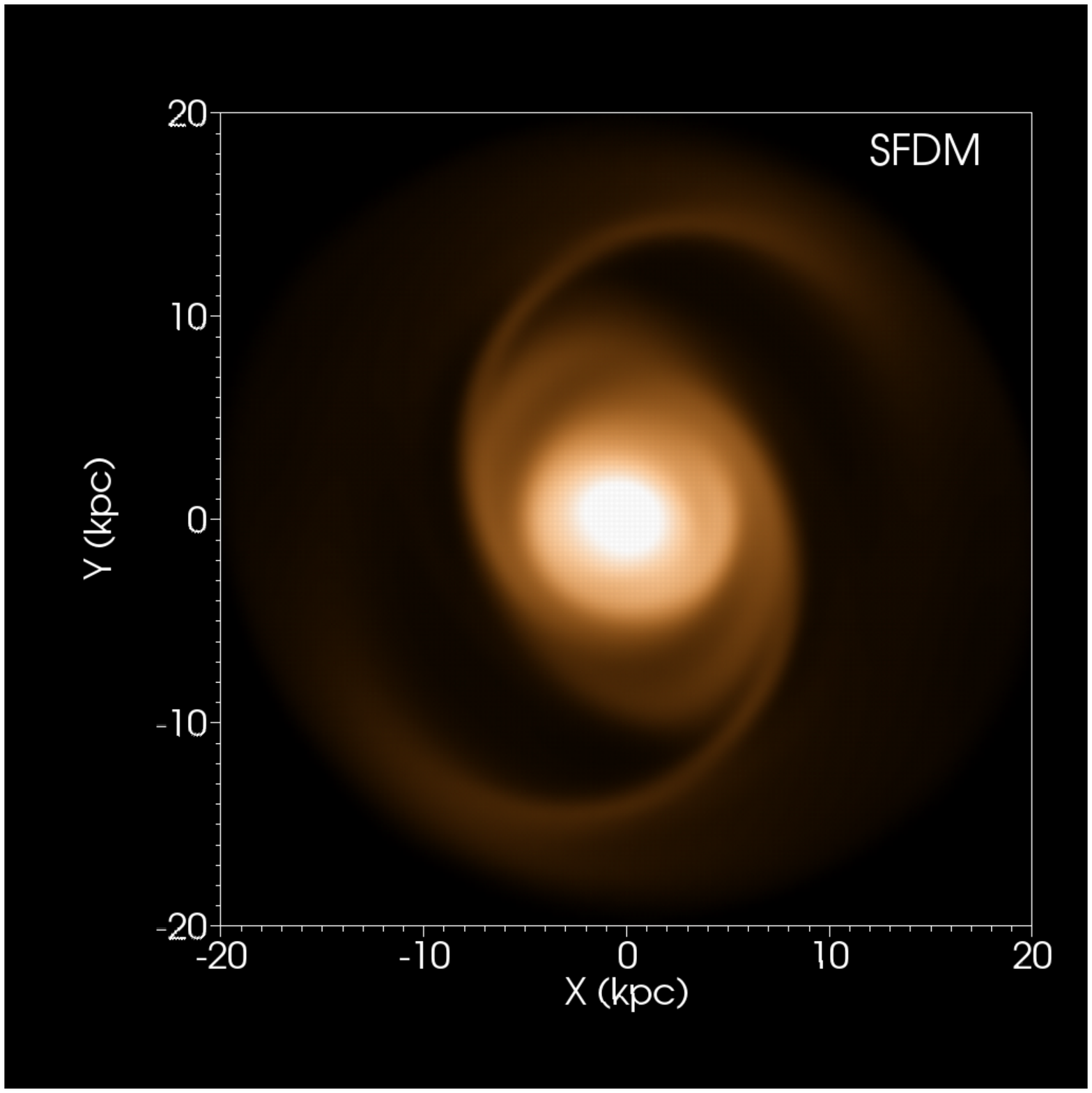}
   \hspace{-0.6cm}   \includegraphics[height=80mm]{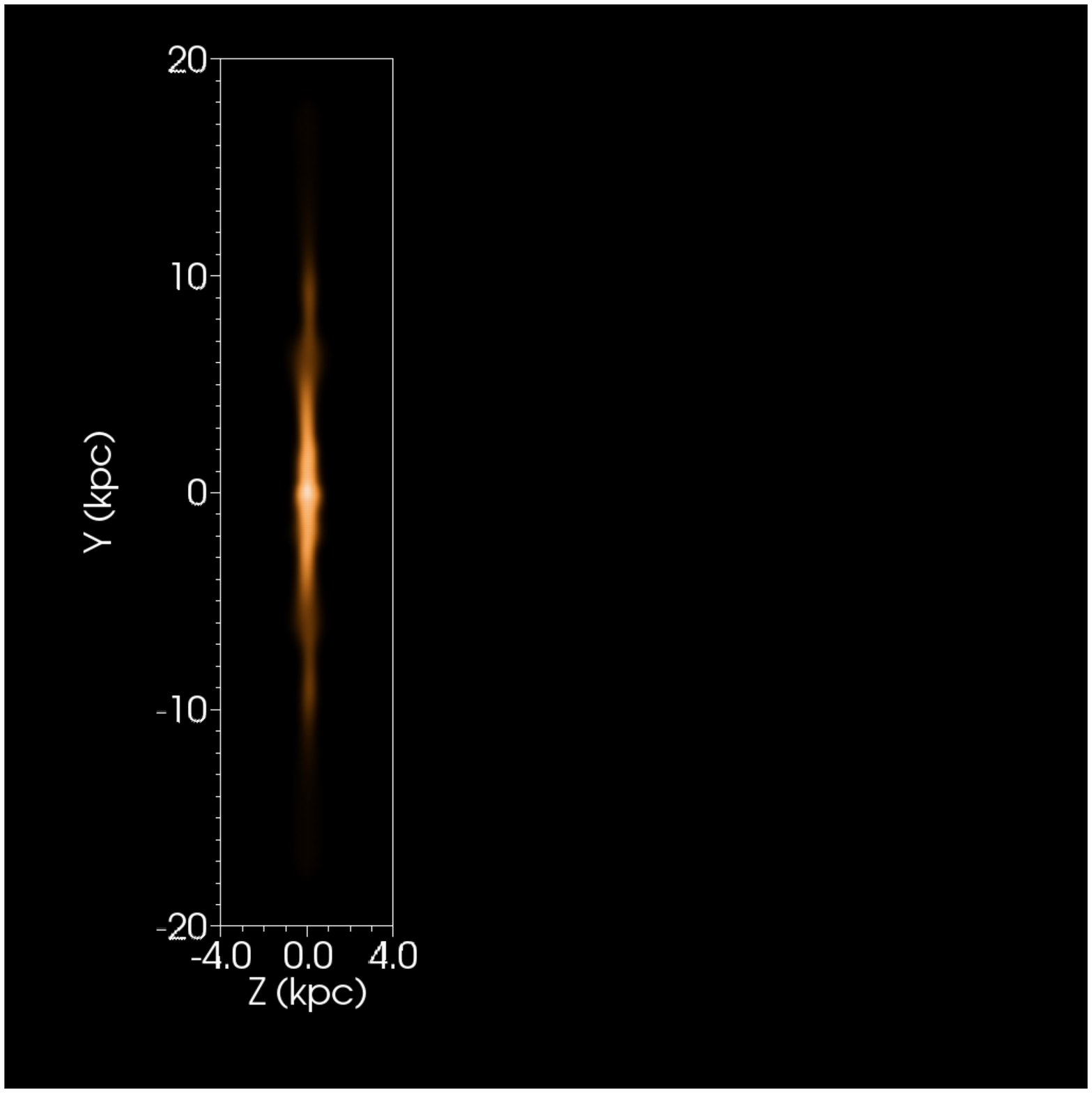} 
   \hspace{-4.3cm} \includegraphics[height=80mm]{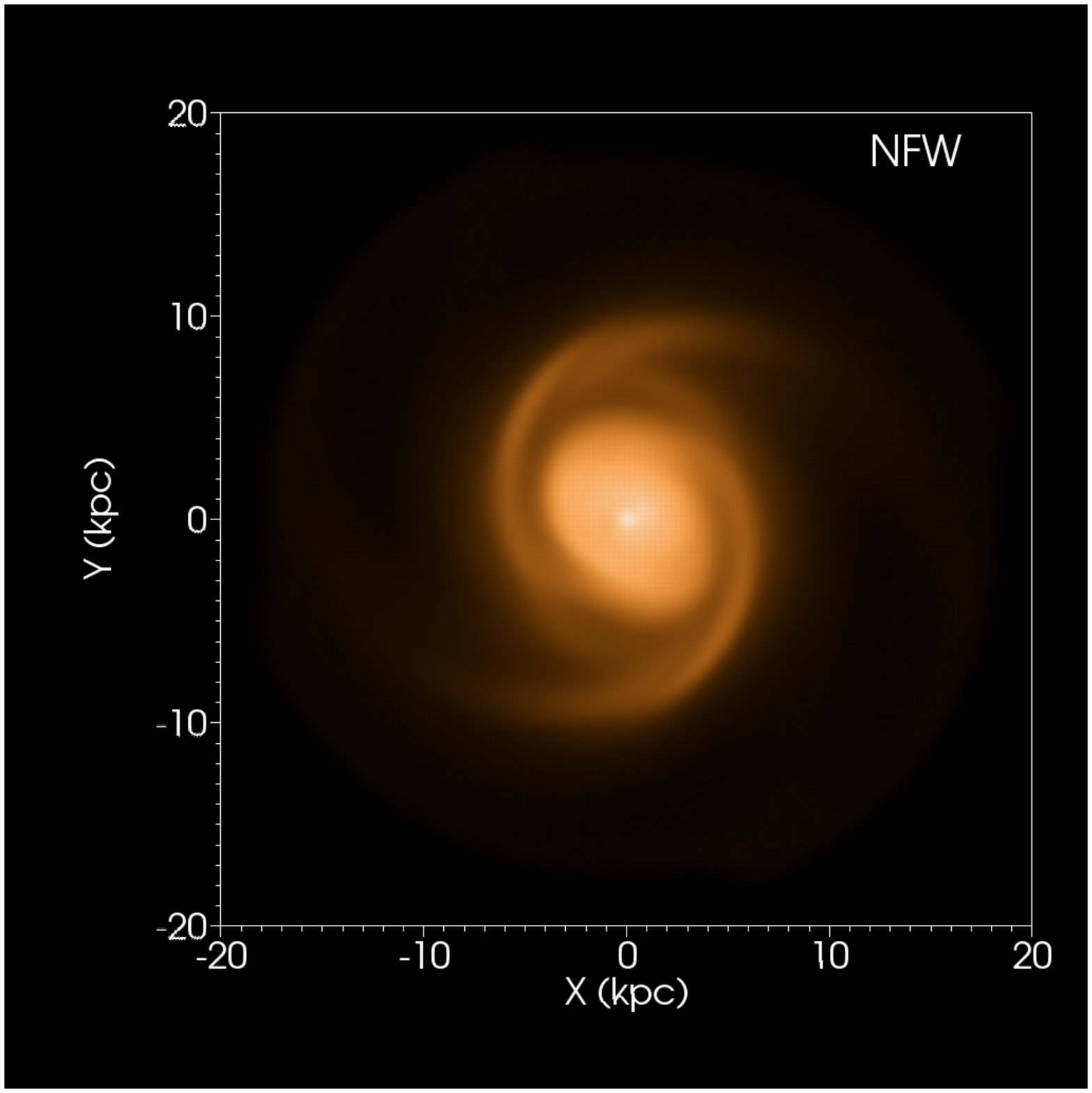}
   \end{center}
   \vspace{-1cm}
   \caption{Face-on and edge-on projections of the gas distributions embedded in a WDM (or SFDM) triaxial halo and a NFW triaxial halo. In both cases the rotating triaxial potentials trigger the formation of spiral patterns.} 
   \label{fig:Simulation1}
\end{figure*}

Figure \ref{fig:Simulation1} shows the results of our simulations. The luminous matter responds to the non-axisymmetric background, and although the triaxial shape in the halos is not achieved in the same way, is able to trigger the formation of a spiral pattern in both cases.

In the particular case of WDM it is an important outcome, that triaxial halos arise naturally, a feature at galactic scales that WDM shares with CDM. This links the model in a natural way with the presence of spiral and barred patterns.

Although the imprints of both halos on the gas are qualitatively the same, is worth trying to differentiate the underlying dark matter distribution, even if both are triaxial.

To this purpose, since having the hydrodynamic simulations allow us to measure directly the rotational velocity of the gas, we outline the rotation curves over perpendicular directions to have an appreciation of the rotation of the gas and its possible dependence with the direction.

\begin{figure*}
   \begin{center}
   \hspace{-1.3cm}   \includegraphics[height=80mm]{SFDM1yzB.pdf} 
   \hspace{-4.3cm} \includegraphics[height=80mm]{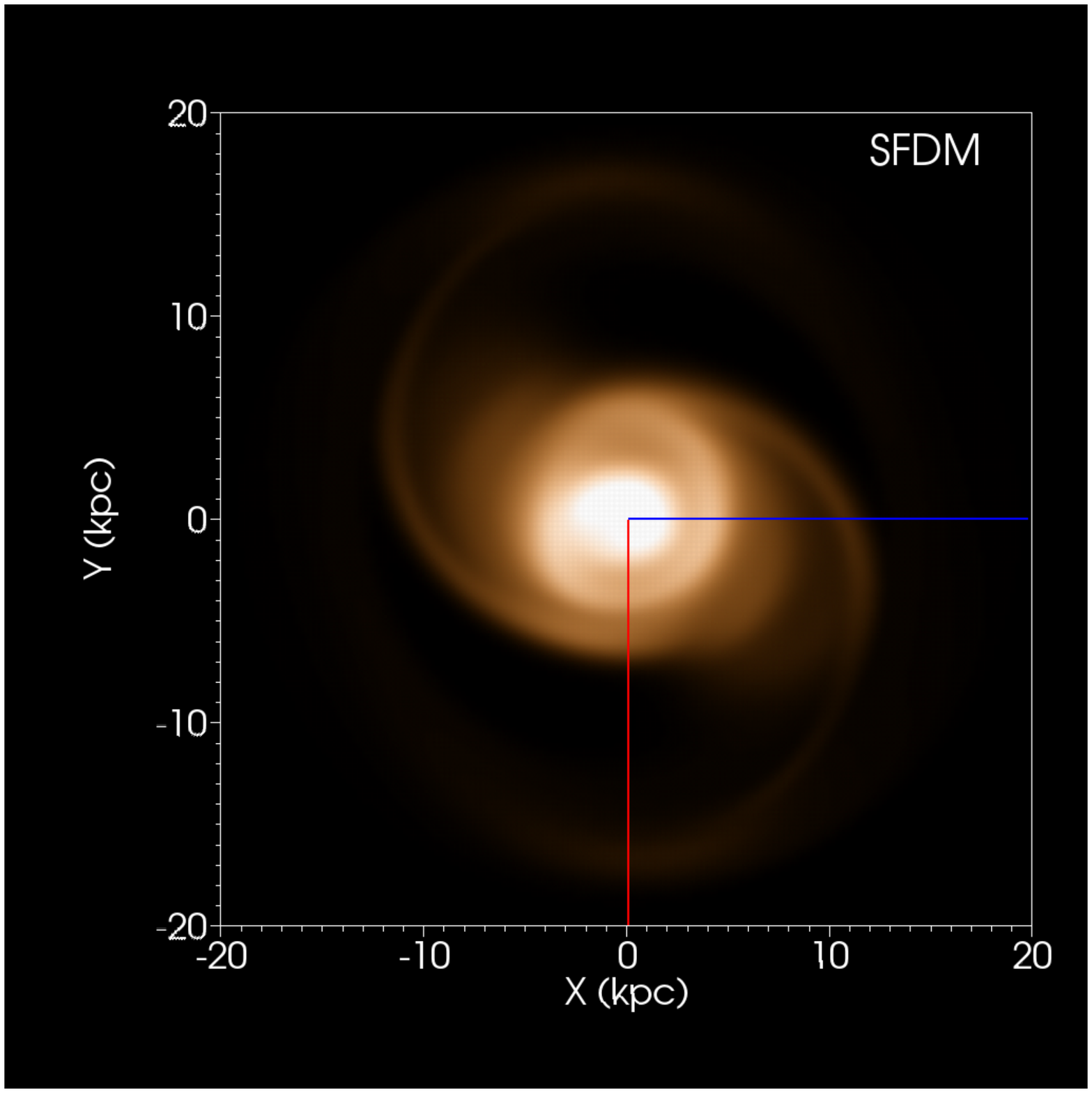}
 \hspace{-0.6cm} \includegraphics[height=80mm]{NFW1yzB.pdf} 
   \hspace{-4.3cm} \includegraphics[height=80mm]{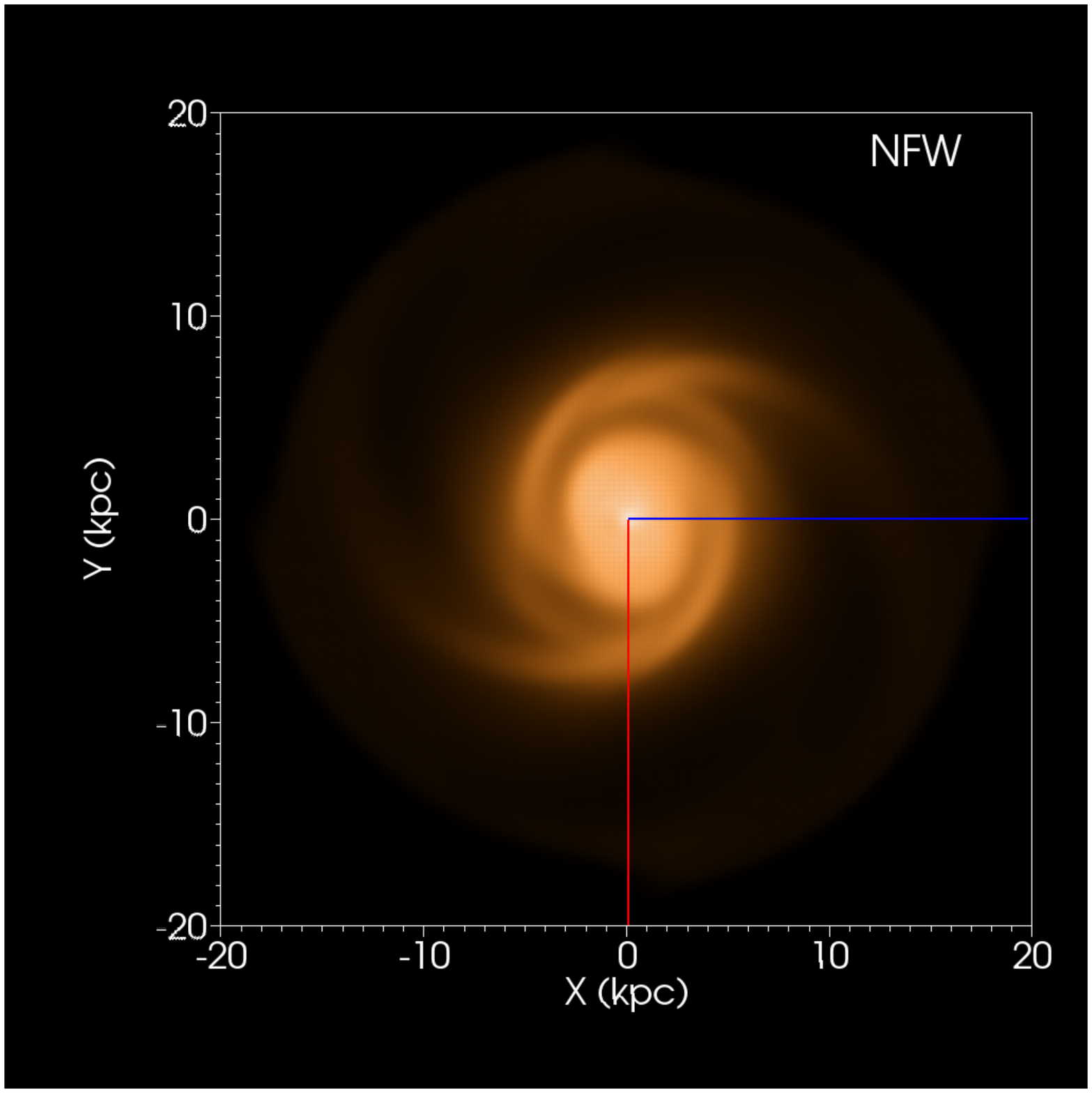} \\
   
   \hspace{-1.8cm} \includegraphics[height=70mm]{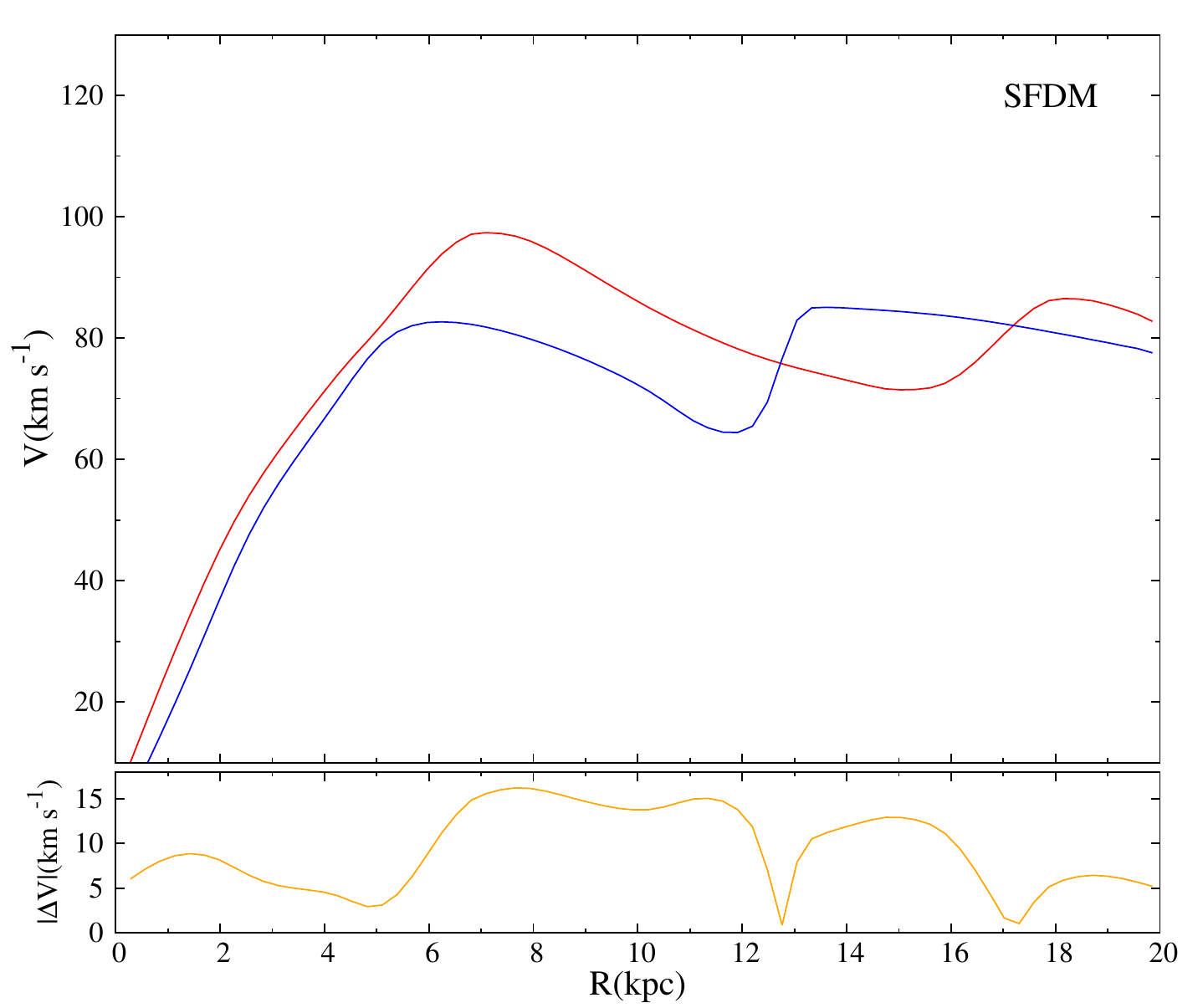}
   \hspace{-0.2cm} \includegraphics[height=70mm]{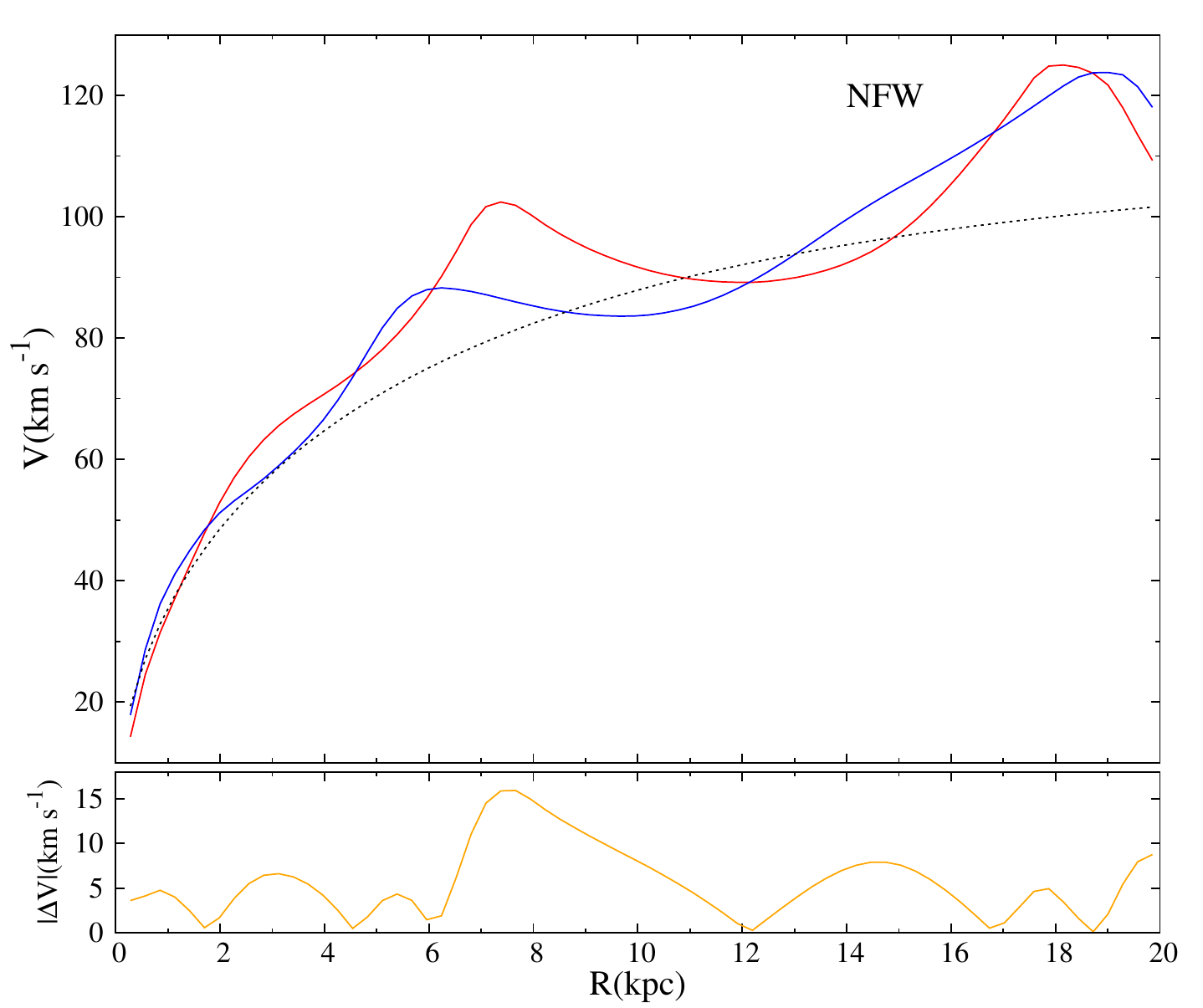}
   \end{center}
   \caption{(Top) Projected gas density, red and blue lines indicate the direction in which we measure the rotation curve. (Bottom) Rotation curves in two different directions for each halo.} 
   \label{fig:RCs}
\end{figure*}

Figure \ref{fig:RCs} shows the rotation curves of the gas embedded in the WDM (or SFDM) and NFW halos
We measure the curves at a time in the simulation where non circular motions are evidently an important part of the gas kinematics. Blue and red lines indicate the directions in which we outline these curves
First we notice that the rotation curves in Figure \ref{fig:RCs} (bottom panel) change with the measurement direction, this changes are more notorious at large radii while in the central region seems to be smaller.

On the bottom of those same plots we quantify this change by computing the difference between the two curves for each halo.

In the case of the WDM halo we can see that $\Delta V$ is nearly constant in the inner $4kpc$, which means that although the curves are different, they rise with the same slope.

For the NFW halo we see a similar behavior, $\Delta V$ is small and nearly constant within the inner $2kpc$. Even more, for the central region the rotation curve along the two directions do not deviate significantly from the circular velocity (dotted line).

The similar behaviour that the rotation curves have in the inner region of both halos is an indication that in spite the triaxiality of the background potential and the evident presence of non circular motions on the gas, observations could still be able to distinguish the central distribution of dark matter.

The relevance of this result for the case developed here and originally in Bray (2010) is that the natural triaxiality of WDM halos preserve the inherent features of a scalar field configuration, a density core that becomes important when trying to explain the observed dark matter distribution in dwarf galaxies. The spiral imprints that the triaxial halos have on the luminous matter do not prevent the models of being tested by observations.

%\begin{figure}
%   \begin{center}
%   \includegraphics[height=100mm]{SFDM1.pdf} \hspace{-0.9cm} \includegraphics[height=100mm]{NFW1.pdf}
%   \end{center}
%   \caption{}. 
%   \label{fig:Simulation1}
%\end{figure}

%%%Once the system has evolved and the gas distribution is no longer axisymmetric, we can measure the rotation curve directly over the gas as shown in Fig. \ref{fig:RC1}. As we can see, even in the presence of non-circular motions, the two systems are distinguishable mostly in their inner regions where the observational data favour the cored WDM halo. 

%\begin{figure}
%   \begin{center}
%   \includegraphics[height=60mm]{RC1.pdf}
%   \end{center}
%   \caption{Rotation curves profiles after the baryonic component has evolved within each of the triaxial halos. Here we compare the curves with the observational data of the LSB galaxy F568-3.}
%   \label{fig:RC1}
%\end{figure}

\begin{figure}
   \begin{center}
   \includegraphics[height=100mm]{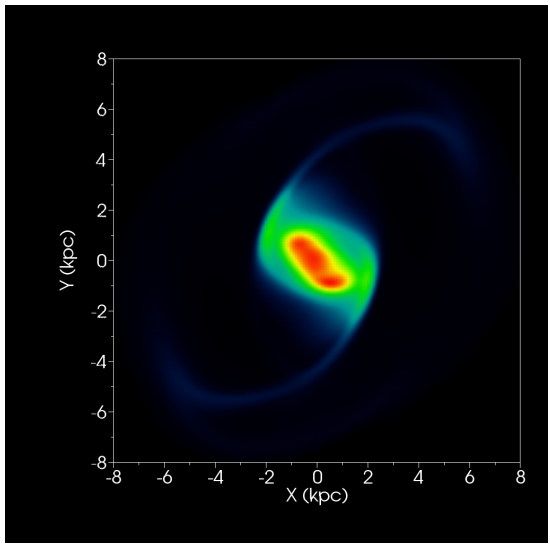}
   \end{center}
   \vspace{-1cm}
   \caption{Face-on projection of the gas density embedded in a smaller, less massive, and less rounded halo whose parameters are adjusted to model the LSB galaxy F583-4. The gas develops open spiral arms and a bar-like structure.
}
   \label{fig:RC2}
\end{figure}

Finally we also adjust the model to describe the properties of a smaller and less massive LSB galaxy, the parameters in eq. \ref{potentialsphericalharmonics} allow us to variate the triaxiality of the dark halo. In Figure \ref{fig:RC2} (top panel) we can see that the imprints of this less rounded halo on the baryonic matter is to stretch the gas distribution, which develops spiral arms with a greater pitch angle (more open arms) compared with the previous simulation. Also the disk central region, initially axisymmetric, develops a bar-like distribution.

%%%Figure \ref{fig:RC2} (bottom panel) shows the rotation curves at two different times for this simulation. Its important to notice that the triaxial shape of the halo, that induce non-circular motions on the gas, implies that the profiles of the rotation curves variate appreciably, mostly in the outer parts. But most important, within the inner 1kpc the rotation curves are unchanged, maintaining a behavior inherent to a cored dark halo in good agreement with the observational data in spite the presence of non-axisymmetric structures and non-circular motions.

\section{Discussion and Conclusions}
\label{sec:conclu}

A large number of observations in LSB and dSph galaxies are better described by dark matter halos with cored density profiles, opposite to the predictions of N-body simulations with collisionless dark matter. A vast number of numerical studies have shown that this discrepancy has a plausible solution in the baryonic feedback which efficiently injects energy to the dark matter particles and therefore flattens the halo center. However the energy requirements and efficiencies for this mechanism to transform cusps into cores are not typical and mostly works at high redshifts \citep{Penarrubia2012, Amorisco2014}. Also, simulations of dwarf galaxies show that even when the feedback produces a core in the center of the halo, a cusp could regrow by the infall of substructures to the galaxy \citep{Laporte2014}.

Because the baryonic feedback solution seems to work only under specific circumstances and do not explains the presence of a core at all galactic scales, it is important to explore alternative DM models that predict cored density profiles.

As in previous studies, here we explore the WDM model but generalizing the density profiles to rotating triaxial halos as this would be a more natural scenario. This is an important outcome for the model, that in general the halos have triaxial shape, equivalent to the halos predicted by CDM simulations.

The fact that the gravitational potential is no longer spherically symmetric or axisymmetric has important effects on the baryonic matter of the galaxy. It is not surprising to see the development of a spiral pattern in the galactic plane, which seems to be a usual outcome for a triaxial potential as shown in our comparison with a triaxial NFW halo.

This is important because when trying to obtain information about the dark matter distribution in a galaxy, the mere presence of spiral arms or a bar may indicate that the baryonic matter dominates the central region and therefore observations like rotation curves may not tell us precisely what the DM distribution is at the halo center. But here we show that spiral arms and bars can be developed in a DM dominated galaxy with a central density core without supposing its origin on mechanisms intrinsic to the baryonic matter. These non-axisymmetric features are well explained and triggered by the interaction of the disk with the WDM triaxial halo, a triaxiality that arise naturally within the model.

\acknowledgments

The authors wish to thank to B. Pichardo, O. Valenzuela and V. Robles for many helpful discussions. The numerical computations were carried out in the ``Laboratorio de Super-C\'omputo Astrof\'{\i}sico (LaSumA) del Cinvestav.''
The authors acknowledge to the General Coordination of Information and 
Communications Technologies (CGSTIC) at CINVESTAV for providing HPC 
resources on the Hybrid Cluster Supercomputer ``Xiuhcoatl.''
This work was partially supported by CONACyT M\'exico under grants CB-2009-01, no. 132400, CB-2011, no. 166212,  and I0101/131/07 C-234/07 of the Instituto Avanzado de Cosmologia (IAC) collaboration
(http://www.iac.edu.mx/). LAMM is supported by a CONACYT scholarship.
HLB gratefully acknowledges support for this research by NSF grant DMS-1406396.

\end{document}